\documentclass[twocolumn,showpacs,preprintnumbers,amsmath,amssymb,floatfix]{revtex4}

\usepackage{graphicx}
\usepackage{epsfig}
\usepackage{float}
\usepackage{psfrag}

\begin{document}

\title{Steady States of a Nonequilibrium Lattice Gas}
\author{Edward Lyman}
\affiliation{Center for Computational Biology and Bioinformatics, University 
of Pittsburgh, Pittsburgh, PA 15206}
\author{B. Schmittmann}
\affiliation{Center for Stochastic Processes in Science and Engineering and 
Department of Physics, Virginia Tech, Blacksburg, VA 24061-0435}
\date{\today}

\begin{abstract}
We present a Monte Carlo study of a lattice gas driven out of equilibrium by
a local hopping bias. Sites can be empty or occupied by one of two types of
particles, which are distinguished by their response to the hopping bias.
All particles interact via excluded volume and a nearest-neighbor attractive
force. The main result is a phase diagram with three phases: a homogeneous
phase, and two distinct ordered phases. Continuous boundaries separate the
homogeneous phase from the ordered phases, and a first-order line separates
the two ordered phases. The three lines merge in a nonequilibrium bicritical
point.

\end{abstract}
\maketitle

\section{\protect\smallskip Introduction}

The statistical treatment of systems driven far from equilibrium presents
exciting theoretical challenges \cite{SZ,MD}. Lacking the unified
understanding afforded equilibrium phenomena by the work of Boltzmann and
Gibbs, we are exploring unknown territory without recourse to an established
theory. Our physical intuition, developed in the context of equilibrium
systems, can be misleading when faced with nonequilibrium problems. We
therefore turn our attention to computational `experiments' in which a
manifestly nonequilibrium state can be established and studied, seeking to
identify key features shared by many nonequilibrium systems. While real
physical systems are unquestionably important, complications and subtle
details may obscure such common features. While motivated by realistic
problems, microscopic rules and boundary conditions are chosen simple enough
to facilitate a comprehensive computational study of the full parameter
space. Many of the questions relevant in equilibrium remain interesting,
especially as concern the nature of phase transitions and the principle of
universality. The richness of nonequilibrium phenomena is often surprising,
as the relaxation of the detailed balance constraint allows a variety of
unexpected possibilities: in contrast to equilibrium, the \emph{dynamics}
now affects the stationary (long-time) properties of the system.
Particularly dramatic effects have been observed in models where the
violation of detailed balance is combined with spatial anisotropies and
dynamic conservation laws \cite{SZ}. There, \emph{effective} long-range
interactions can be induced even if the microscopic rules are perfectly
local in space and time \cite{eff-LR}.

In this paper, we consider a model from this class, namely, a lattice gas of
two species of particles and holes on a fully periodic lattice in two
spatial dimensions. To drive the system out of equilibrium, we bias the
hopping rates of the two species in opposite directions, reminiscent of an
`electric' field, $E$, acting on opposite `charges' (though we stress that
there is \emph{no Coulomb interaction}). A nonzero charge current signals
the nonequilibrium steady state. The two species interact through an
excluded volume constraint and nearest-neighbor attractions. We choose the
interactions carefully, in order to unify three important models which
appear as limiting cases of our more general theory. First, by letting all
particles attract each other, irrespective of their identity (charge), the
non-driven limit corresponds to the familiar Ising lattice gas \cite{Ising}.
This well-known equilibrium model will serve as an anchor for our studies of
driven systems. Turning the bias on but removing all members of one species,
we recover the \emph{driven }Ising lattice gas introduced by Katz et al \cite%
{KLS} (the KLS model). The third limit, obtained by letting the interaction
strength vanish, corresponds to a non-interacting two-species model first
proposed by Schmittmann et al \cite{SHZ} (the SHZ model). While the KLS\
model phase-separates, via a continuous transition, into high- and
low-density strips \emph{aligned} with the field \cite{KLS}, the SHZ model
orders into \emph{transverse, }charge-\emph{\ }and mass-segregated, strips %
\cite{SHZ}, similar to jamming instabilities in traffic models \cite{traffic}%
. If a charge imbalance is imposed, these strips drift \cite{LZ}. Our study
will allow us to bridge the gap between these very different scenarios. To
set the scene, we briefly survey some of the relevant previous work.

As a single parameter modification of the Ising lattice gas, the KLS model
is a \emph{minimal model} for the study of nonequilibrium steady states
(NESS). Particle hops along one direction (parallel to the $x$-axis) occur
at the normal equilibrium rate, as if in contact with a heat bath at
temperature $T$. Particle hops in the other direction (parallel to the $y$%
-axis) are enhanced (suppressed) in the positive (negative) direction by
coupling to an external field, $E$. With periodic boundaries in the $y$%
-direction, a nonzero current is maintained, and the system settles into a
NESS. At half-filling, there remains a continuous transition, though with $%
T_{c}\left( E\right) $ increasing monotonically with $E$ and saturating at $%
1.414T_{c}\left( E=0\right) $ \cite{ItalDDS,LeungDDS,WangDDS}. The
transition falls into a novel universality class with exponents distinct
from the Ising ones \cite{FTDDS}. The critical behavior is strongly
anisotropic, with distinct sets of exponents characterizing fluctuations 
\emph{perpendicular} and \emph{parallel }to $E$. There has been some
discussion regarding the nature of these fluctuations, with some authors
disputing the original claim that the correct mesoscopic description is
gaussian in the perpendicular direction \cite{debate}. Though the anisotropy
makes numerical investigations of the critical behavior quite subtle and
computationally intensive, recent high precision Monte Carlo studies compare
the two mesoscopic descriptions. The results are in complete agreement with
the predictions of the original field theory \cite{ItalDDS}. As a final
note, we mention that the combination of anisotropic dynamics and a
conservation law introduces power law correlations at all $T>T_{c}$ \cite%
{eff-LR}, a manifestation of the relaxation of the detailed balance
constraint. These correlations are revealed by the structure factor, which
has a discontinuity singularity at the origin. In this sense, even the
`disordered' phase is quite non-trivial.

Turning to multispecies versions, the simplest (`SHZ') model \cite{SHZ}
allows \emph{two} different types of particles, distinguished only by their
interaction with the external field. Positive (negative) particles are
biased to hop in the positive (negative) $y$-direction, and interact only
via an excluded volume constraint. The temperature is absorbed into $E$ and
the only parameters are $E$, the overall mass density $m$, and the overall
charge density (i.e., the density difference of the two species), $f$. Here,
the mechanism for ordering is the mutual volume exclusion of the particles,
so that at sufficiently strong $E$ and large $m$, the system locks into a
high density strip \emph{perpendicular} to $E$, with positive and negative
particles blocking each other. For non-zero charge density $f$, this strip
is found to drift in the direction of the minority species \cite{LZ}.
Depending on where the phase boundary is crossed, first-order or continuous
transitions are observed \cite{VZS,Korniss}. Various other remarkable
properties have been discovered. For a range of aspect ratios,
configurations with non-zero winding number (`barber poles') are quite
frequently observed, \emph{in addition} to the usual transverse strips,
raising the possibility of bistability \cite{BSZ}. Power law correlations
characterize the disordered phase, with \emph{directionality-dependent}
exponents \cite{2sp-corr}. Another subtle issue concerns the lower critical
dimension: While an exact solution for a strictly one-dimensional model,
characterized by a single `lane' parallel to $E$, precludes a transition %
\cite{mukamel}, Monte Carlo data for a `two-lane' model indicate the
presence of a macroscopic cluster in finite systems \cite{2xL}. Very
subtle finite-size effects control the decay of this cluster in the 
thermodynamic limit \cite{Kafri,Athens,Georgiev}.

In this paper, we consider the two-species model at finite $T$ and $E$,
where interparticle interactions are expected to play an important role. By
varying $T$, $E$, and $f$, the fraction of the \emph{total} population which
are of the \emph{minority} species, we can interpolate smoothly from the KLS
model to the (non-interacting) two-species SHZ model. Hence, we expect a
competition between the two types of ordered configurations -- parallel vs
transverse strips -- favored by these two limits. As $f$ varies from $0.0$
(KLS model) to $0.5$ (equal numbers of each), there should be some critical 
$f$ where the preferred order switches. To explore these phenomena in more
detail, we map out the phase diagram in $E,$ $f$ and $T$, for a range of
system sizes. The energy scale is set by our choice of the interparticle
attraction $J$, and the overall mass density $m$ is fixed at $0.5$ so that
the Ising critical point remains accessible. Many questions arise in
connection with earlier work. How do nearest-neighbor
attractions modify the two-species transition? What will be the effect of a
few `impurities' (i.e., minority particles) on the KLS transition? At what
concentration do the `impurities' become relevant and change the nature of
the transition? Preliminary results, focusing on a restricted parameter
space, were already reported in \cite{LS}; here, we explore a much wider
parameter range, including several system sizes. 
We will be able, if not to answer these questions fully,
then to at least suggest the character of their resolution sufficiently to
guide further research. 

Our main results are as follows. At fixed $E$ and sufficiently small $f$, a
line of continuous transitions emerges from the pure KLS ($f=0.0$) point,
in the $f$-$T$ plane.
This line separates the disordered phase from an ordered one, characterized 
by a particle-rich strip parallel to $E$. 
As we increase $f$, we encounter a bicritical
point, where the transition line splits into a line of continuous
order-disorder transitions, from disorder into a strip transverse to $E$, 
and a line of first-order transitions along which transverse and parallel 
order coexist.
If we fix $f$ and lower $T$, we first observe the transition from disorder
into the transverse strip, followed by a transition into parallel order. 
This topology persists at higher $E$, except that all lines are shifted to
slightly higher temperatures. The size-dependence of the phase diagram is
subtle, since the main features are controlled by \emph{different} scaling
variables. On the one hand, the transition into the transverse strip is
controlled by the \emph{effective} drive $L_{y}E/T$ where $L_{y}$ is the
system size in the drive direction. On the other hand, the bicritical point
appears to depend on the scaling variable $L_{y}f$ which translates into the
number of \emph{rows} (transverse to $E$) which can be filled with the
minority species. Finally, the pure KLS point requires finite-size scaling
at fixed shape factor $A\propto L_{y}/L_{x}^{3}$ \cite{LeungDDS}, in two
spatial dimensions. 

The remainder of the paper is organized as follows. We first describe in
detail the microscopic model and the observables which are used to locate
the different phases. We then present our simulation results, beginning with
the structure of typical configurations in different parts of parameter
space and their associated order parameters. By monitoring the signatures
of first and second order transitions, we compile a cut through the phase
diagram at fixed $E$, with variable $f$ and $T$. The phase boundaries and
their dependence on system size are analyzed in some detail. To complete the
picture, we present two cuts at different but fixed temperatures, crossing
the phase boundaries by varying $E$ and $f$. We conclude with a brief
summary and a discussion of some open questions.

\section{ Microscopic Model and Observables.}

We consider periodic square lattices of size $L_{x}\times L_{y}$, in two
spatial dimensions, with $E$ parallel to the (positive) $y$-axis. A
configuration is specified by the set of occupation variables, $\left\{
\sigma \left( \mathbf{r}\right) \right\} $, where $\sigma \left( \mathbf{r}%
\right) $ takes three values, $\pm 1,0$ denoting a positive (negative)
particle or a hole at lattice site $\mathbf{r}$. Often, we will only need to
distinguish particles from holes, via $n\left( \mathbf{r}\right) \equiv
\left| \sigma \left( \mathbf{r}\right) \right| $. All lattices are half
filled, i.e., $m\equiv \left( L_{x}L_{y}\right) ^{-1}\sum_{\mathbf{r}%
}n\left( \mathbf{r}\right) =1/2$, so that the Ising critical point remains
accessible. An important parameter is the fraction of negative particles
(the `minority species') in the system: $f=\left( mL_{x}L_{y}\right)
^{-1}\sum_{\mathbf{r}}\delta _{-1,\sigma \left( \mathbf{r}\right) }$.
Clearly, we only need to consider the sector $0\leq f\leq 0.5$, from having
no negative particles at all to equal numbers of each species. For later
reference, we also introduce the charge density, $q\equiv \left(
L_{x}L_{y}\right) ^{-1}\sum_{\mathbf{r}}\sigma \left( \mathbf{r}\right) =m-f$. 
The nearest-neighbor attraction is modelled by the Ising Hamiltonian 
\begin{equation}
H=-4J\sum\limits_{\left\langle \mathbf{r,r}^{\prime }\right\rangle }n\left( 
\mathbf{r}\right) n\left( \mathbf{r}^{\prime }\right)  \label{Is-Ham}
\end{equation}%
%
%
We choose attractive interactions, $J>0$, regardless of species. While many
other choices are possible and interesting, ours provides maximum linkage to
known cases: Ising, KLS and SHZ. The Monte Carlo dynamics conserves the
number of each species and is specified as follows. An update attempt begins
by picking a bond at random. If the bond connects a particle-hole pair, the
contents are exchanged with the Metropolis rate $\min \left\{ 1,\exp \left[
-\left( \Delta H-\delta yE\sigma \left( \mathbf{r}\right) \right) /T\right]
\right\} $ \cite{Metrop}. Hence, at $E=0$ we recover the equilibrium Ising
model with conserved magnetization, coupled to a heat bath at temperature $T$. 
We will set $J=1$ and measure $E$ in units of this (arbitrary) energy
scale. $T$ will be quoted in units of the Onsager temperature, $T_{c}(E=0)$.
The change in $y$-coordinate, due to the proposed move, is denoted by 
$\delta y$, and $\Delta H$ is the associated change in internal energy. The
term $\delta yE\sigma \left( \mathbf{r}\right) $ models the gain or loss of
energy from the coupling to $E$; if $\delta y\sigma \left( \mathbf{r}\right) 
$ is positive (negative) the move is favored (unfavored). Our model, in
which $E$ and $T$ are varied independently, raises an interesting issue. If
the ratio $E/T$ is quite large, it becomes almost impossible for particles
to hop backwards. In a finite system, this implies that a relatively small
fraction of the minority species -- provided the `right' fluctuation occurs
-- is sufficient to form a stable blockage. Even though such a fluctuation
becomes less probable in a larger system, the dynamics nevertheless becomes
nonergodic in the limit $E/T\rightarrow \infty $. In principle, this can be
avoided by introducing, e.g., a small probability for particles to exchange
places \cite{Korniss}. To limit the number of parameters, we circumvent
these problems here by considering different initial configurations and a
range of system sizes.

The dynamics is diffusive, and therefore conserves both charge and mass
density. Though the \emph{local} effect of the external field is analogous
to the effect of an electrostatic potential on electric charges, the
boundary conditions exclude the possibility of a \emph{global} Hamiltonian
description.

As overall density is conserved, we expect ordered configurations to be
strips of higher density coexisting with strips of lower density. We
therefore introduce the Fourier transform of the local mass variable, 
\begin{equation}
\tilde{s}(m_{x},m_{y})\equiv \frac{\pi }{L_{x}L_{y}}\sum_{x,y}n(x,y)e^{2\pi
i(m_{x}x/L_{x}+m_{y}y/L_{y})}  \label{ft}
\end{equation}%
which is labelled by (integer) wavenumbers, $m_{x}=0,1,...,L_{x}$, 
$m_{y}=0,1,...,L_{y}-1$. The structure factor, 
\begin{equation}
S(m_{x},m_{y})\equiv \left\langle \left| \tilde{s}(m_{x},m_{y})\right|
^{2}\right\rangle   \label{S}
\end{equation}
%
then serves as a good order parameter, since it is sensitive to \emph{%
mass-segregated }strip configurations. For example $S\left( 1,0\right) $
will be $\mathcal{O}\left( 1\right) $ for a strip aligned with the field,
characteristic of KLS order; similarly, $S\left( 0,1\right) $ will detect a
strip transverse to $E$ which develops in the SHZ (two-species) model; and
both are normalized to $\mathcal{O}\left( 1/L_{x}L_{y}\right) $ for a disordered
configuration. We also monitor a `susceptibility', i.e. the fluctuations of
the order parameter: 
\begin{eqnarray}
\Delta (m_{x},m_{y})\equiv L_{x}L_{y}\left[ \left\langle \left| \tilde{s}%
(m_{x},m_{y})\right| ^{4}\right\rangle \right. \label{Delta}\\
\left. - \left\langle \left| \tilde{s}%
(m_{x},m_{y})\right| ^{2}\right\rangle ^{2}\right]  \nonumber
\end{eqnarray}
%
We note that $S(m_{x},m_{y})$ involves the Fourier transform of the \emph{%
mass} variable and is therefore not sensitive to any charge-segregated
structures. Replacing $n(x,y)$ by $\sigma (x,y)$ in Eq.~(\ref{S}) generates
structure factors which respond to charge inhomogeneities. We have monitored
these and their fluctuations throughout, and found that their behavior is
consistent with the mass-based quantities.

When $S$ is calculated, the average is taken over multiple steady-state
configurations of a Monte Carlo run, with a typical run lasting $0.8$M 
($8 \times 10^5$) Monte
Carlo steps (MCS), and $2L_{x}L_{y}$ bond update attempts per MCS. Data are
collected every $400$ MCS; fluctuations of observables indicate that this
interval is sufficient to produce uncorrelated data in the largest $\left(
60\times 80\right) $ systems considered. Typically, the initial $0.2$M MCS
are discarded to ensure that data are taken from the steady state. Near
critical points and at low temperatures these numbers require modification,
due to long correlation times and long-lived metastable states. In such
cases the only recourse is a careful analysis of individual, very long runs.
When that is necessary we will measure a quantity closely related to $S$: 
\begin{equation}
s(m_{x},m_{y})\equiv \left| \tilde{s}(m_{x},m_{y})\right| ^{2}  \label{tt}
\end{equation}%
%
%
which measures the type of order present in a \emph{single} configuration.
We can then track $s$ for different $m_{x}$'s and $m_{y}$'s over the course
of a run and see precisely how the averages are generated.

Finally, at each bond update we tally the quantity $\delta y\sigma \left(
x,y\right) $, which is then averaged over a run to give the charge current $%
j $. However, it is not particularly illuminating to compare $j$ for
different values of the effective drive $E/T$. We are more interested in a
quantity which is a property of the gross structure of the steady state,
namely, the conductivity $\kappa $ defined via $j=\kappa E/T$.

\begin{figure}[tp]
\begin{center}
\vspace{-0.1cm}
\begin{minipage}{\columnwidth}
  \epsfig{file=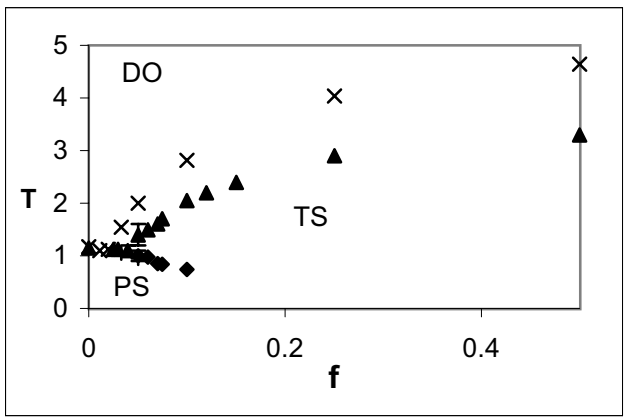, width=\columnwidth}
    \vspace{-1.0cm}
\end{minipage}
\end{center}
\caption{Phase diagram in $f$ and $T$ for $E=2.0$. Triangles and diamonds
are boundaries in the $40\times40$ system, $\times$'s are for the $60\times
60$ system.} 
\vspace{-0.2cm}
\label{fig1}
\end{figure}
Now that we have described the various quantities which will be used to
probe the behavior of our model, we turn to the presentation of the data.

\section{Results}

\subsection{Phase diagram in $f$ and $T$}

In this section we seek the location and character of transitions by
scanning in $f$ and $T$ at fixed drive $E$. We choose $E=2.0$ since this
intermediate value still allows for a significant fraction of backward
jumps, thus avoiding the spurious metastable configurations discussed above.
At the same time, it is large enough to induce measurable currents and other
clear signatures of far-from-equilibrium behavior. The two order parameters $%
S\left( 1,0\right) $ and $S\left( 0,1\right) $ and their fluctuations are
monitored in order to identify the different phases. Large peaks in their
fluctuations, or the presence of hysteresis, are used as indicators of
continuous vs first order transitions, respectively. For clarity, we first
present a quick overview of the topology of the phase diagram, and then turn
to the details of the data which underlie this picture.

\begin{figure}[tp]
\begin{center}
\vspace{-.1cm}
\begin{minipage}{\columnwidth}
  \epsfig{file=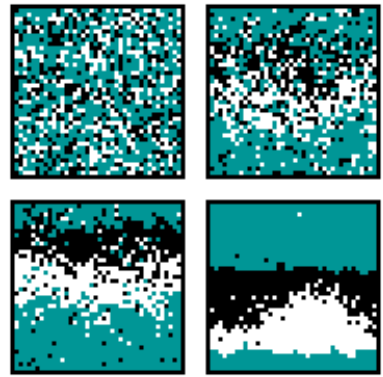, width=.6\columnwidth}
    \vspace{-.4cm}
\end{minipage}
\end{center}
\caption{Configurations of the $40\times40$ lattice at $f=0.50$ for four
temperatures. Upper left, $T=6.0$; upper right, $T=3.25$; lower left, $T=2.64$;
lower right, $T=1.0$. Positive (negative) particles are white (black); $E$
points upwards.} 
\vspace{-0.5cm}
\label{fig3}
\end{figure}
Fig.~\ref{fig1} shows the phase diagram in the $f$-$T$ plane, at $E=2.0$,
for two different system sizes. Three phases are found: a homogeneous,
disordered phase (DO), a transverse strip (TS) phase as in the two-species
model, and a parallel strip (PS) phase as in the KLS model. The value of $f$
determines which phase is observed: at $f=0$ there is only one species of
particles, and a single transition is observed from disorder into the
parallel strip. As $f$ increases, this transition persists until the number
of the minority species is sufficient to create a blockage and form the
transverse strip. From here on, \emph{two} transitions are observed: from
disorder into the transverse strip, and at a lower temperature from the
transverse strip into the parallel strip. Upon increasing $f$ further, only
the DO-TS transition can be detected. Although the TS-PS line cannot have a
critical endpoint for reasons of symmetry, for $f>0.10$ it occurs at such a
low temperature that it cannot be observed in simulations of a reasonable
length.

Now that we have briefly discussed the phase diagram we can look in more
detail at the phases and their boundaries. Before presenting the data
(structure factors, their fluctuations, currents) we will begin with some
pictures of typical configurations at various points in the phase diagram.
This way the reader can develop some intuition about the model. All pictures
show $40\times 40$ lattices, with $E$ pointing up. White (black) pixels are
positively (negatively) biased particles, blue-green (gray in print) pixels
are holes. We will begin at the right side of the phase diagram and explore
how typical configurations change as we move left, decreasing the fraction
of the population which belongs to the minority species.

\begin{figure}[tp]
\begin{center}
\vspace{.1cm}
\begin{minipage}{\columnwidth}
  \epsfig{file=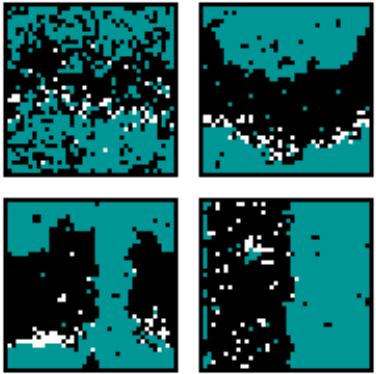, width=.6\columnwidth}
    \vspace{-.4cm}
\end{minipage}
\end{center}
\caption{Configurations of the $40\times40$ lattice at $f=0.075$ for four
temperatures. Upper left, $T=1.77$; upper right, $T=0.95$; lower left, 
$T=0.84$; lower right, $T=0.78$.} 
\vspace{-0.5cm}
\label{fig5}
\end{figure}
In Fig.~\ref{fig3} we present four configurations at $f=0.50$: The first
picture is at $T=6.0$, well above $T_{c}=3.3$. Unsurprisingly, this
configuration lacks any visible structure since it falls deep into the
homogeneous phase. This will be the only picture presented for the
homogeneous phase, as the only \emph{visible} feature which changes with $f$
is the ratio of white to black pixels. The next picture is at $T=3.25$, and
now we begin to see the two-species type phase separation. Lowering the
temperature further to $T=2.64$, the horizontal strip appears quite clearly,
though it remains diffuse at the boundaries, and there are many `travelers'
(isolated particles) moving through the remainder of the system. At this
temperature backward hops of particles are not too improbable, occuring with
a rate $\exp (-E/T)=0.47$, and thus a fair number of holes are able to enter
the strip and allow particles to slip through the blockage. Finally at $%
T=1.0 $ we are deeply inside the ordered phase, and the boundaries of the
strip are very sharp and travelers are few. At this temperature, holes are
rarely able to penetrate into the interior of the strip. The irregular shape
of the interface between the two species is a result of the quench from
disorder; a more gradual lowering of the temperature would result in a
smoother interface.

Lowering $f$ to $0.075$ (Fig.~\ref{fig5}), only $1.5$ rows of the minority
species are left. The first frame shows a configuration just below
criticality at $T=1.77$. Though a blockage can still form, the strip is no
longer symmetric with respect to $+$ and $-$. This leads to a drifting of
the strip: Occasionally, the rather thin blockage of the minority species is
opened by backward hops, and the majority species pours through. These
particles then travel quite rapidly around the periodic lattice and attach
to the back of the majority blockage, the net result being an \emph{upward}
drift of the strip. Lowering $T$ further to $0.84$ (third frame) it appears
that interfaces \emph{parallel} to $E$ are becoming favorable; this type of
configuration is common at these intermediate values of $f$: here, parallel
and transverse strips compete with each other. Indeed the final frame ($%
T=0.78$) shows the preferred low temperature configuration: a single strip
of \emph{mixed} charge parallel to $E$, raising the possibility of a
sequence of \emph{two} transitions as a function of $T$.

Now that we have developed a qualitative notion of the various regions in
the phase diagram, we turn to a more quantitative analysis of the phases and
their boundaries. As in the preceding section, we present the order
parameters and their fluctuations as a function of $T$, for a range of $f$,
though with one important difference: we will show data for at least \emph{%
two} system sizes. While this cannot replace a high-precision finite-size
scaling analysis of the transitions, it is intended to provide a rough
picture of how the fluctuations of the order parameter scale with system
size.
\begin{figure}[tp]
\begin{center}
\vspace{-.1cm}
\begin{minipage}{\columnwidth}
  \epsfig{file=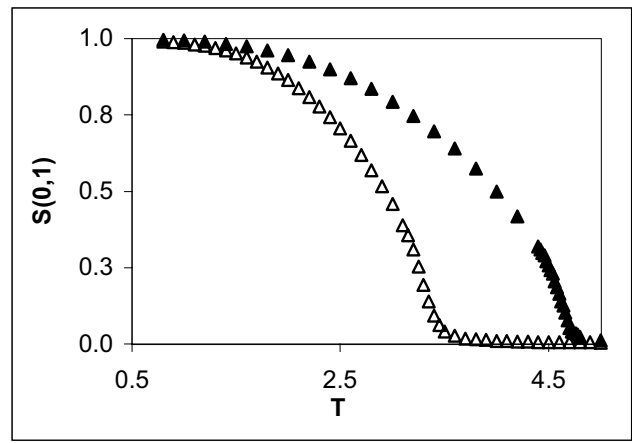, width=\columnwidth}
    \vspace{-.9cm}
\end{minipage}
\end{center}
\caption{$S\left(0,1\right)$ as a function of $T$ for $f=0.50$. Open (filled)
triangles are for the $40\times40$ ($60\times60$) system.} 
\vspace{-.2cm}
\label{fig10}
\end{figure}

\begin{figure}[tp]
\begin{center}
\vspace{-.1cm}
\begin{minipage}{\columnwidth}
  \epsfig{file=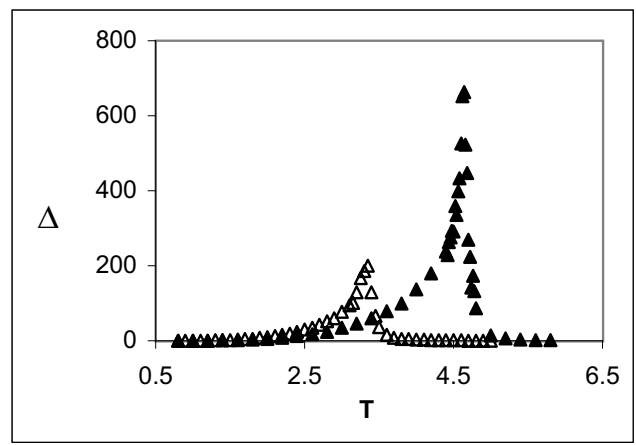, width=\columnwidth}
    \vspace{-.9cm}
\end{minipage}
\end{center}
\caption{$\Delta\left(0,1\right)$ as a function of $T$ for $f=0.50$. Open
(filled) triangles are for the $40\times40$ ($60\times60$) system.} 
\vspace{-.2cm}
\label{fig11}
\end{figure}

We begin as before at $f=0.50$, with equal numbers of each type of particle.
In Figs.~\ref{fig10} and~\ref{fig11}, $S(0,1)$ and its fluctuations are
plotted as a function of $T$ for two system sizes ($40\times 40$ and $%
60\times 60$); $S(1,0)$ is not shown as the only transition here is from the
homogeneous phase into the transverse strip. In both systems, $S(0,1)$ goes
smoothly to zero as $T$ is increased. A clean peak in $\Delta (0,1)$ is also
observed in each system, increasing in amplitude with the system size. These
two observations are consistent with a continuous transition into the
transverse strip at $f=0.50$, and we therefore use the location of the peak
in $\Delta (0,1)$ to locate the phase boundary in Fig.~1, with $T_{c}\left(
L=40\right) =3.35$ and $T_{c}\left( L=60\right) =4.64$.

Reducing $f$ to just a few rows of the minority species introduces some new
features into the data, which we describe briefly before continuing
our analysis. Here, we still observe the transition described above, with
the same behavior of $S(0,1)$ and $\Delta (0,1)$. However, now that the
strip is no longer symmetric, it drifts. 
The ordered phase therefore fluctuates significantly, and $\Delta
(0,1)$ develops a shoulder at about $1/3$ of the value it reaches at the
transition. We will see below that the fluctuations of the ordered phase can
obscure the transition in smaller systems.

\begin{figure}[bp]
\begin{center}
\vspace{-.5cm}
\begin{minipage}{\columnwidth}
  \epsfig{file=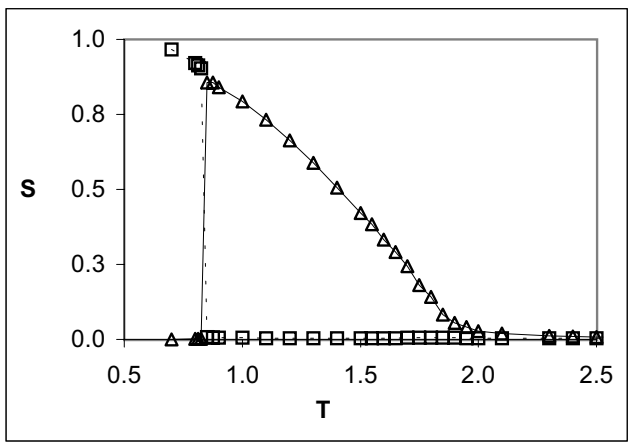, width=\columnwidth}
    \vspace{-.9cm}
\end{minipage}
\end{center}
\caption{$S\left(0,1\right)$ and $S\left(1,0\right)$ as a function of $T$
for $f=0.075$ in the $40\times40$ system. Triangles (squares) are for 
$S\left(0,1\right)$ ($S\left(1,0\right)$). Lines and dashes are provided
to guide the eye.} 
\vspace{-.1cm}
\label{fig14}
\end{figure}
\begin{figure}[tp]
\begin{center}
\vspace{-.1cm}
\begin{minipage}{\columnwidth}
  \epsfig{file=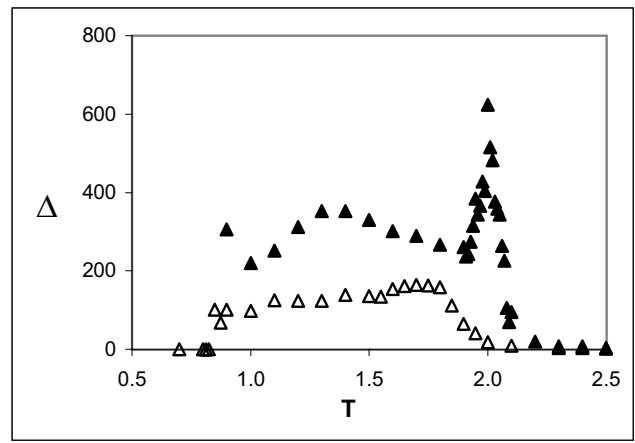, width=\columnwidth}
    \vspace{-.9cm}
\end{minipage}
\end{center}
\caption{$\Delta\left(0,1\right)$ as a function of $T$ for $f=0.075$. Open
(filled) triangles are for the $40\times40$ ($60\times60$)system.} 
\vspace{-.2cm}
\label{fig15}
\end{figure}
\begin{figure*}[t]
\begin{center}
\vspace{-.1cm}
\begin{minipage}{\textwidth}
  \epsfig{file=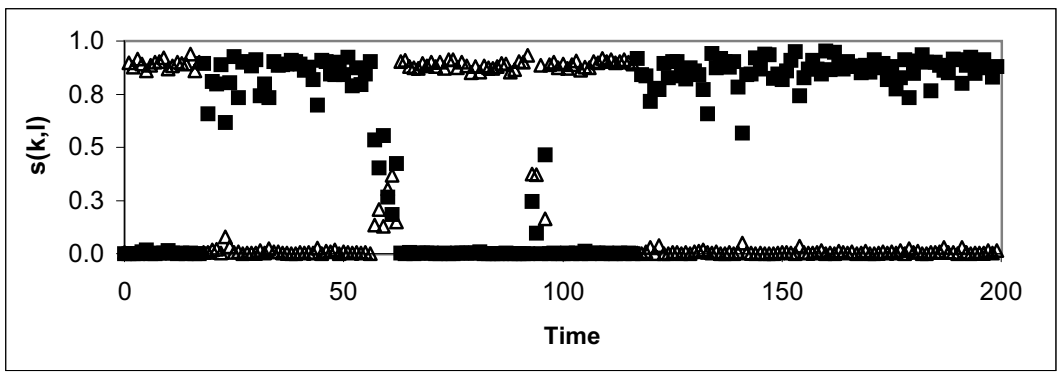, width=.8\textwidth}
    \vspace{-.5cm}
\end{minipage}
\end{center}
\caption{Timetrace at $T=0.832$. Time in units of $0.2$M MCS is plotted on the
horizontal axis. The values of $s\left(0,1\right)$ and $s\left(1,0\right)$
(triangles and squares, respectively) are plotted on the vertical axis.} 
\vspace{-.1cm}
\label{fig16}
\end{figure*}

Upon reducing $f$ further we observe a sequence of \emph{two transitions} as
a function of $T$. Fig.~\ref{fig14} shows both order parameters, $S(0,1)$
and $S(1,0)$ for $1.5$ rows of the minority species: $f=0.075$ in the 
$40\times 40$ system. Also shown in Fig.~\ref{fig15} is $\Delta (0,1)$ for 
$1.5$ rows of the minority species in both the $40\times 40$ and the 
$60\times 60$ system. We have omitted the $60\times 60$ data for the order
parameters to keep the plot uncluttered. As before, $S(0,1)$ and $\Delta
(0,1)$ signal a continuous transition into the TS phase as $T$ is
lowered, though the signal in $\Delta (0,1)$ is much more pronounced in the 
$60\times 60$ system. And also as before, there are significant fluctuations
associated with this phase, due to strip drifting. Specifically, in the 
$60\times 60$ system $\Delta (0,1)$ actually has a broad secondary peak in
the ordered phase. The large magnitude of this signal is is quite unexpected 
and awaits a satisfactory explanation. For now, we only note that 
\emph{lowering} $T$ \emph{increases} the effective
bias, $E/T$ , and therefore enhances fluctuations associated with the drive.
Lowering $T$ further we observe $S(0,1)$ falling abruptly, while $S(1,0)$
climbs rapidly, suggesting a discontinuous transition from the horizontal
strip (TS) into the vertical strip (PS). In the neighborhood of such a transition, 
one expects to see metastability of the unfavored phase, and this is indeed the
case as shown in Fig.~\ref{fig16}. Here we have plotted \emph{time traces}
(as opposed to configurational averages) of structure factors for individual
configurations, $s\left( 1,0\right)$ and $s\left( 0,1\right)$, defined
in Eqn.~(\ref{tt}). Clearly, when $s\left( 1,0\right)$ ($s\left( 0,1\right)$) 
$=1$ the configuration is a perfect vertical (horizontal) strip.
Sufficiently close to the transition, the time traces reveal the expected
behavior, as the system switches between the two ordered phases. Notice the
length of the run shown: $40$M MCS, which is a factor of $40$ longer than
typical runs, indicating that the lifetimes of metastable configurations are
already quite long even in the $40\times 40$ system, rendering such behavior
nearly unobservable in the $60\times 60$ system.

At smaller values of $f$ we are nearing the junction of the three phase
boundaries, which considerably complicates the analysis of data from small
systems in a couple of ways. The sequence of transitions (DO-TS followed by
TS-PS) becomes difficult to resolve, as they are quite close in temperature,
and massive fluctuations from the first-order TS-PS line may wash out the
signal in $\Delta (0,1)$ which locates the continuous DO-TS line. And if the
junction of the three lines is indeed a nonequilibrium bicritical point, we
can expect unexplored finite-size effects to interfere with the analysis. We
can, however, make some progress based on the assumption that the relevant
control parameter near the bicritical point is the number of rows of the
minority species. This hypothesis will be treated in more detail below, in
the sections on scaling arguments.

At precisely one row of the minority species it is no longer possible to
accurately resolve the two transitions in the $40\times 40$ system. A weak
transverse ordering is observed, with $S(0,1)$ reaching at most $40\%$ of
perfect order. In the vicinity of the PS-TS first-order transition, huge
fluctuations associated with switching between metastable configurations are
observed, which wash out the signal of the DO-PS transition. However, it is
interesting that these transitions can be resolved in larger systems at
precisely one row of the minority species: There, these two transitions are
sufficiently far apart in temperature since (as we will see below) $T_{c}$
increases with $L_{y}$ across the DO-TS transition. This likely explains why
the DO-TS transition is not observed in the $40\times 40$ system.
\begin{figure}[bp]
\begin{center}
\vspace{-.5cm}
\begin{minipage}{\columnwidth}
  \epsfig{file=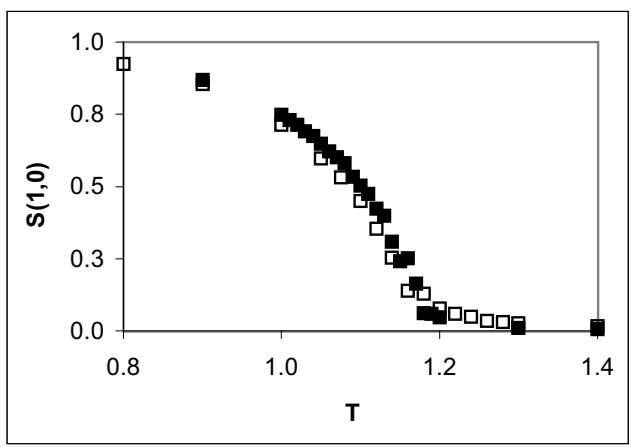, width=\columnwidth}
    \vspace{-.9cm}
\end{minipage}
\end{center}
\caption{$S\left(1,0\right)$ as a function of $T$ for $f=0.025$. $40\times40$
($60\times60$) data are shown by open (filled) squares.} 
\vspace{-.1cm}
\label{fig25}
\end{figure}

Below one row of the minority species we no longer observe the transverse
order, though we caution that this may be strictly correct only for the 
\emph{finite} system. With $f$ just below a single row of the minority
species, the high temperature phase is homogeneous and the low temperature
phase is the parallel strip. In the vicinity of the transition, huge
fluctuations are observed in $S\left( 0,1\right) $ and especially $S\left(
1,0\right) $, where the fluctuations are an order of magnitude larger than
the signal at the DO-TS boundary. In this region neither $S\left( 1,0\right) 
$ nor $S\left( 0,1\right) $ posesses a well-defined average; timetraces
indicate vigorous competition between the two ordered phases. We conjecture
that we are close to the bicritical point in the finite system, and are
therefore unable to resolve the transition without some knowledge of scaling
to guide the analysis. At smaller $f$ we are farther from the bicritical
point, and the complications from the presence of the minority species are
less severe. Fig.~\ref{fig25} shows $S\left( 1,0\right) $ for two system
sizes at exactly one-half row of the minority species. We have not shown 
$S\left( 0,1\right) $, as the signal has become insignificant. $S\left(
1,0\right) $ shows the low-temperature configuration to be a single parallel
strip, the smooth approach to zero again suggesting a continuous transition.
And indeed, the data for $\Delta \left( 1,0\right) $ is consistent with this
conjecture, with a sharp peak which increases with system size. The
amplitude of the peak is smaller by a factor of five than the peaks observed
near the bicritical point, suggesting that it is associated with
fluctuations about a well-defined average, as opposed to transitions into
and out of the ordered phase. Timetraces support this conjecture.

\begin{figure}[bp]
\begin{center}
\vspace{-.5cm}
\begin{minipage}{\columnwidth}
  \epsfig{file=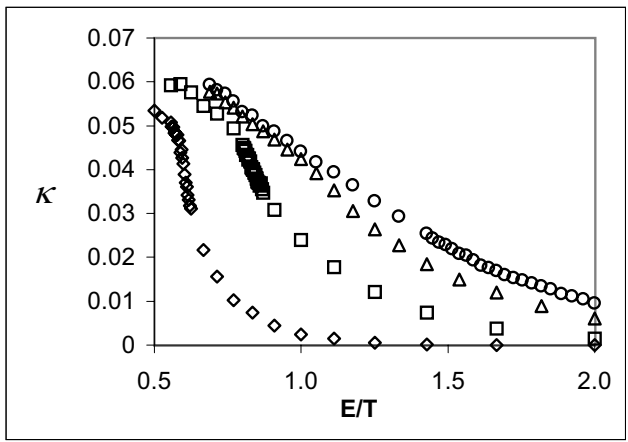, width=\columnwidth}
    \vspace{-.9cm}
\end{minipage}
\end{center}
\caption{Conductivity as a function of effective drive $E/T$ for several $f$. 
$f=0.50$, diamonds; $f=0.10$, squares; $f=0.075$, triangles; $f=0.05$,
circles; $f=0.04$, $+$'s. Lines have been added to the largest and smallest 
systems to guide the eye.} 
\vspace{-.1cm}
\label{fig28}
\end{figure}

Charge currents were also measured in order to look for signatures of the
various phases and transitions. We note however that as we are changing $T$,
we are also changing the \emph{effective} bias $\varepsilon \equiv E/T$, and
therefore a more appropriate quantity is the conductivity, $\kappa \equiv
j/\varepsilon $. In Fig.~\ref{fig28} we plot $\kappa $ as a function of 
$\varepsilon $ for each $f$ discussed in this and the following sections; we
present only data for the $40\times 40$ system as there are no significant
differences between the two system sizes. At temperatures below $T=1.4$ ($%
\varepsilon >1.4$) the conductivity vanishes in the $f=0.50$ systems; as the
temperature is raised the effective drive is reduced, and backward hops
occasionally occur, allowing a small current to trickle through the
blockage. Upon raising the temperature further, the conductivity changes
slope at $T=3.6$ in the $40\times 40$ system, which \emph{does not} coincide
with the phase transition. Rather the maximal conductivity occurs in the
disordered phase, but at a temperature which is not too large, so that
backward hops are not so common as to begin reducing the current. The
transition apparently corresponds instead to the \emph{inflection point} ($%
T=3.4$), where the curvature changes sign. At $f=0.10$, the conductivity has
a slope discontinuity in the $40\times 40$ system at $T=2.6$, slightly above
the critical temperature $T_{c}=2.4$. Though there should be an inflection
point in $\kappa $ near the transition, our data are not precise enough to
locate it. At $f=0.075$ and below, the conductivity drops smoothly to zero
with $T$, showing no indication of either transition. Apparently, $S$ and $%
\Delta $ are far more sensitive to the transitions in our model.

We close this section on the $f$-$T$ phase diagram with a summary of the
results. The picture at higher $f$ is clear: a clean continuous transition
into the horizontal strip, with $T_{c}$ decreasing with $f$. When $f$ is
reduced to approximately three rows of the minority species, the signal of
the transition remains clear, though it now sits atop a shoulder of
fluctuations of the ordered phase. At yet smaller $f$, a second transition
appears between the two ordered phases at lower $T$; it has the
characteristics of a first-order transition. At even lower $f$, at
approximately one row of the minority species, both fluctuations
perpendicular and parallel to $E$ become so violent that the DO-TS
transition is only seen in larger systems, as the two transitions nearly
overlap in the $40\times 40$ system. Close to the $f=0$ point, the
transition is once again clean and apparently continuous into a vertical
strip of mixed charge. Of course, all these statements are based on an
analysis of \emph{finite} systems. In order to draw robust conclusions, a
more systematic analysis of larger samples is required.

\subsection{Phase diagram in $f$ and $E$}

In the preceding section we studied a slice of the phase diagram at constant 
$E$, varying the fraction of the minority species and the temperature.
Varying $T$ effectively varies both the strength of particle-particle
attractions \emph{and} the strength of the bias, since the relevant
quantities in the rates are $J/T$ and $E/T$. In this section we consider a
different cut through the phase diagram. By varying $E$ and $f$ at fixed $T$%
, the interparticle attractions are held constant while the strength of the
bias is varied. In this way we can study directly the competition between
the drive and the attractive interactions. In the following, we choose a
value for $T$, and then scan in $E$ for several values of $f$. The
temperatures are chosen by reference to the KLS temperature: $T=2.0$ is
above the critical temperature of the KLS model at saturation, and $T=1.2$
is at the critical temperature for $E=2.0$, studied in the previous section.
At this stage, we have only data for $40\times 40$ systems, and are
therefore as yet unable to speculate on results for larger systems. However,
they cast a new light on the more detailed results of the previous sections.
\begin{figure}[tp]
\begin{center}
\vspace{-.3cm}
\begin{minipage}{\columnwidth}
  \epsfig{file=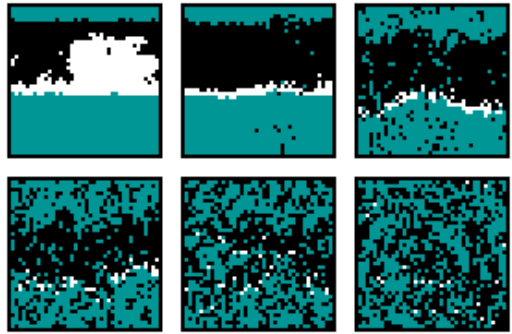, width=.8\columnwidth}
    \vspace{-.5cm}
\end{minipage}
\end{center}
\caption{Configurations for several $f$ at $E=20.0$, $T=2.0$.
Upper left, $f=0.50$; upper middle, $f=0.10$; upper right, $f=0.075$; 
lower left, $f=0.05$; lower middle, $f=0.04$; lower right, $f=0.025$.} 
\vspace{-.2cm}
\label{fig36}
\end{figure}

\subsubsection{$T = 2.0$}

As before, we first survey the phase diagram with the help of some typical
configurations. The qualitative picture will then be made more quantitative
in the next section, by examining the behavior of currents and order
parameters.

Fig.~\ref{fig36} shows a series of configurations at various $f$ for $E=20.0$%
. The first frame clearly shows the transverse strip at $f=0.50$, and the
absence of travelers suggests that the strip is stationary. In the next
frame we have reduced $f$ to $0.10$, reducing the thickness of the minority
species to exactly two rows. Consequently we now see some travelers
trickling through a break in the blockage. Watching an animation in this
region of the phase diagram reveals an interesting behavior: the strip is
mostly quiescent, except for a few particles hopping back and forth at the
particle-hole interface. Aside from the different ratio of $+$ to $-$, these
configurations look similar to the $f=0.50$ strip. Then, a sudden large
fluctuation opens up a hole in the minority blockage: the $+$ particles pour
through, and the strip fluctuates and drifts partway around the lattice,
until the blockage is reestablished. Reducing $f$ further to $0.075$ (third
frame) we see a strip in the middle of one of these fluctuation events.
In contrast to $f=0.10$ where such large fluctuations are
relatively rare, the situation is now reversed; i.e., the quiescent periods
become less frequent. In the fourth frame, we set $f=0.05$, and while the
strip is still clearly visible it now drifts continuously. The final two
frames show $f=0.04$ and $0.025$. Now there is no longer any clear evidence
of phase separation. This rough picture is consistent with our earlier
investigation of the two-species transition, where we observed transverse
order at and above a single row of the minority species.
\begin{figure}[bp]
\begin{center}
\vspace{-.2cm}
\begin{minipage}{\columnwidth}
  \epsfig{file=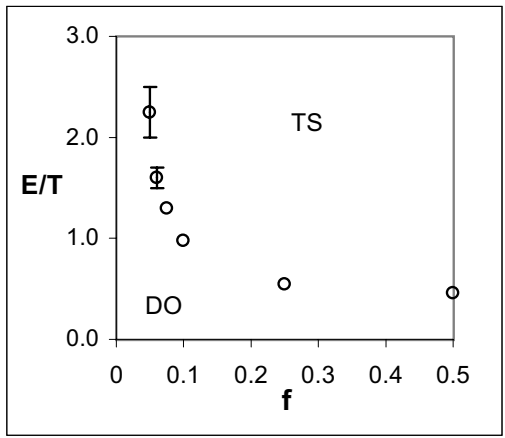, width=.8\columnwidth}
    \vspace{-.6cm}
\end{minipage}
\end{center}
\caption{Phase diagram in $f$ and $E/T$ at $T=2.0$. Disorder (DO) is observed
at small $f$ and $E$, the transverse strip (TS) dominates 
at high $f$ and $E$. The error bars are smaller than the size of the data points,
unless indicated.} 
\vspace{-.2cm}
\label{fig37}
\end{figure}

\begin{figure}[tp]
\begin{center}
\vspace{-.1cm}
\begin{minipage}{\columnwidth}
  \epsfig{file=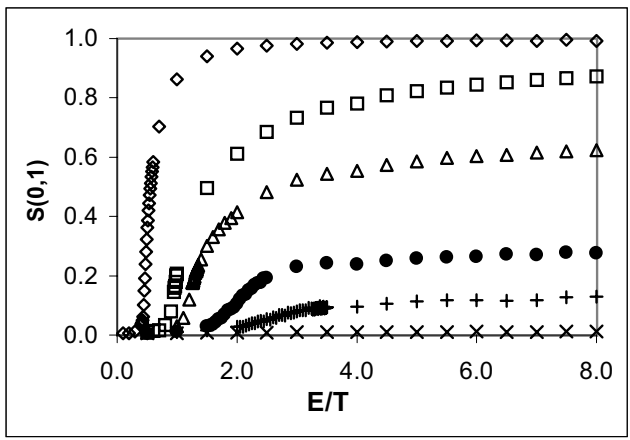, width=\columnwidth}
    \vspace{-.9cm}
\end{minipage}
\end{center}
\caption{$S\left(0,1\right)$ as a function of $E/T$ for several $f$. 
$f=0.50 $, diamonds; $f=0.10$, squares; $f=0.075$, triangles; 
$f=0.05$, circles; $f=0.04$, $+$'s; $f=0.025$, $\times$'s.} 
\vspace{-.2cm}
\label{fig38}
\end{figure}

\begin{figure}[bp]
\begin{center}
\vspace{-.5cm}
\begin{minipage}{\columnwidth}
  \epsfig{file=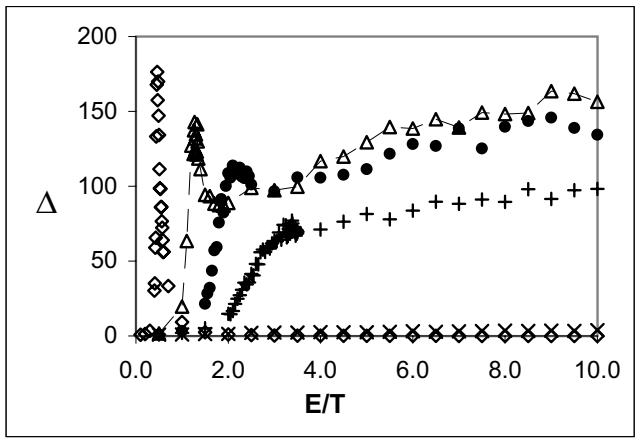, width=\columnwidth}
    \vspace{-.9cm}
\end{minipage}
\end{center}
\caption{$\Delta\left(0,1\right)$ as a function of $E/T$ for several $f$.
The symbols are the same as in the previous figure. 
A line has been added to the $f=0.075$ data to guide the eye.} 
\vspace{-.1cm}
\label{fig39}
\end{figure}

Fig.~\ref{fig37} presents the phase diagram at $T=2.0$. The boundary
separates a horizontal strip at high $E$ and $f$ from a homogeneous phase at
small $E$ and $f$. As we are above the critical temperature for the KLS
model at saturation bias, the vertical strip does not appear at any $E$; at
low $f$ (where we might otherwise expect to see such ordered configurations)
the system simply remains disordered for any $E$ and $f$. Figs.~\ref{fig38}
and~\ref{fig39} show $S\left( 0,1\right) $ and $\Delta \left( 0,1\right) $
for several values of $f$; in the interest of clarity $\Delta \left(
0,1\right) $ is plotted for only four $f$'s. As $E$ is increased the system
orders into a transverse strip, with $S\left( 0,1\right) $ saturating at
smaller values as $f$ is decreased. At $f=0.05$ (exactly one row of the
minority species) $S$ saturates at only $0.28$, indicating that the
transverse ordering is rather weak, though comparison with the data for $%
f=0.025$ shows dramatically different behavior. Here $S\left( 0,1\right) $
reaches a maximum of only $0.01$, and the behavior can hardly be called
`saturation'. $\Delta \left( 0,1\right) $ also signals a transition, though
the clean, sharp spike at $f=0.50$ becomes a broad bump at $f=0.05$, and
shows no signal at $f=0.025$. The susceptibility also indicates a difference
in the ordered phases at different $f$: at large $f$ increasing $E$
suppresses fluctuations, while at smaller $f$ (when the strip begins
drifting) increasing $E$ enhances fluctuations. It is important to note that
the fluctuations at high $E$ are fluctuations about the ordered phase, as 
$\Delta \left( 0,1\right) $ is always $2$ to $3$ orders of magnitude larger 
than $\Delta \left( 1,0\right) $. Though the data is not included in the plots, 
we have checked the behavior of the ordered phase for $E$ as high as 
$40$. The fluctuations for small $f$ ($f<0.10$) saturate and bounce around a
well-defined average, while for larger $f$ they are suppressed. This is true
for $f=0.05$ and greater; at $f=0.025$ the magnitude of fluctuations in
either direction are comparable. We also examined the conductivity
across the phase boundary. We do not present the data here, as it 
is qualitatively similar to the data in Fig.~\ref{fig28}. 
Again, the transition apparently coincides with the inflection
point of the conductivity.

\subsubsection{$T = 1.2$}

\begin{figure}[bp]
\begin{center}
\vspace{-.5cm}
\begin{minipage}{\columnwidth}
  \epsfig{file=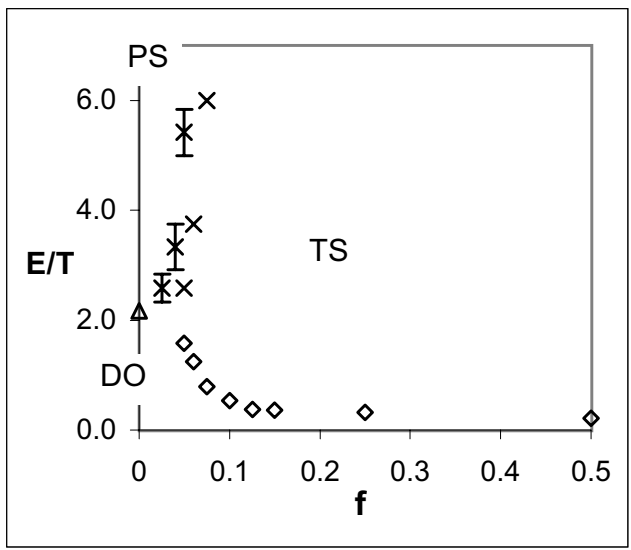, width=\columnwidth}
    \vspace{-.9cm}
\end{minipage}
\end{center}
\caption{Phase diagram in $E/T$ and $f$ for $T=1.2$. A continuous boundary
separates DO from TS (diamonds) as well as DO from PS (triangles). 
$\times$'s indicate a possible boundary between TS and PS. A few typical 
error bars are shown.} 
\vspace{-.1cm}
\label{fig41}
\end{figure}

In the preceding section we found a simple phase diagram, with a single
boundary separating two-species order from the homogeneous phase. Since the
temperature was chosen well above the maximum KLS transition temperature, 
$T_{KLS}\left( E=\infty \right) =1.414$, the KLS ordered phase could not be
observed. Now, we lower the temperature to $T=1.2<T_{KLS}$ and explore the
corresponding ($f$-$E$) slice of the phase diagram (Fig. ~\ref{fig41}). As
long as we remain at $f>0.10$, we observe a transition similar to the one at 
$T=2.0$: from the homogeneous phase into two-species order. In contrast, for 
$f<0.04$ the KLS transition is observed, since the minority species is too
scarce to form a blockage and $T<T_{KLS}\left( E=\infty \right) $. \ Between
these two limiting values of $f$ we are again in the vicinity of the
bicritical point, and the situation becomes complicated due to the competing
types of order.

\begin{figure}[tp]
\begin{center}
\vspace{-.1cm}
\begin{minipage}{\columnwidth}
  \epsfig{file=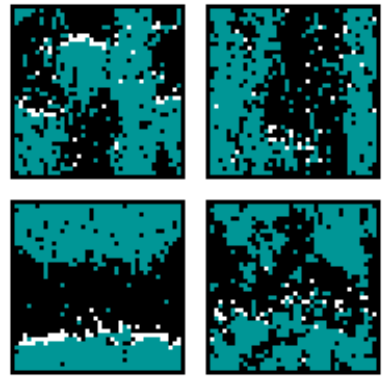, width=.6\columnwidth}
    \vspace{-.5cm}
\end{minipage}
\end{center}
\caption{Typical configurations for the $T=1.2$ plane.
Upper left, $f=0.075$, $E=20.0$; upper right, $f=0.05$, $E=20.0$; 
lower left, $f=0.075$, $E=4.0$; lower right, $f=0.05$, $E=2.4$.} 
\vspace{-.2cm}
\label{fig42}
\end{figure}

Fig.~\ref{fig42} shows typical configurations at two different values of $f$
for various $E$. For $f>0.075$ these configurations look much as they did at
$T=2.0$, so they need not be included. The $f=0.075$, $E=20.0$ configuration
shows some very interesting structure, almost `equal parts' KLS and
two-species order, suggesting some serious competition between the two
phases. We stress that this is a \emph{typical} configuration. When $f$ is
reduced to a single row ($f=0.05$) this competition is reduced, and we see
instead a KLS phase with some local two-species order. This trend continues
upon reducing $f$ to zero, where at high $E$ the KLS order is observed. The
other panels show $f=0.075$ and $0.05$ at smaller values of $E$, $E=4.0$ and 
$2.4$ respectively. These $E$ values were chosen because they maximize the
two-species order for these $f$'s. In each case the strip drifts rapidly
around the lattice. Interestingly, as the majority species is piled onto the
back of the drifting strip it builds long fingers, leading to a very
irregular interface. Pictures at smaller $f$'s are not shown, as they do not
form the transverse strip at smaller $E$. In the next section we locate
these boundaries and study the phases in greater detail by examining the
behavior of $S$, $\Delta $, and $\kappa $.

\begin{figure}[bp]
\begin{center}
\vspace{-.5cm}
\begin{minipage}{\columnwidth}
  \epsfig{file=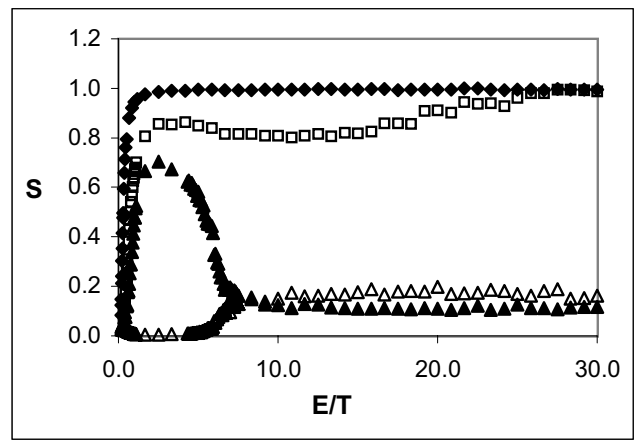, width=\columnwidth}
    \vspace{-.9cm}
\end{minipage}
\end{center}
\caption{$S\left(0,1\right)$ as a function of $E/T$ for several $f$. 
$f=0.50 $, diamonds; $f=0.10$, squares; $f=0.075$, filled triangles. 
Open triangles indicate $S\left(1,0\right)$ for $f=0.075$.} 
\vspace{-.1cm}
\label{fig44}
\end{figure}

In Fig.~\ref{fig44} we plot $S$ for the high-$f$ phases. For $f=0.50$ and
$0.10$, $S\left( 0,1\right) $ shows the system ordering into the
two-species phase much as in the previous section. There is, however, one
notable new feature for $f=0.10$: we see a slight suppression of $%
S\left( 0,1\right) $ over an intermediate range of $E$, from about $E=6$ to 
$E=20$. In this range backward hops occur frequently enough to
occasionally open a hole in the minority blockage, allowing the majority
species to pour through until the blockage reforms. Increasing $E$ reduces
the likelihood of such events, until the strip is perfect except for some
surface diffusion at the particle-hole interface. 
At $f=0.075$ the behavior
changes dramatically. After maximizing the two-species order at $E=3.0$,
increasing $E$ further suppresses $S\left( 0,1\right) $ and enhances 
$S\left( 1,0\right) $, until both saturate below $0.2$. The system can
hardly be said to have switched to KLS order, though neither is it precisely
disordered either. To understand this peculiar behavior, we recall that 
$T$ (and therefore $J/T$) is held constant, so that increasing $E$ 
favors interfaces \emph{parallel} to $E$. Yet, the number of minority 
particles is large enough to form bubbles of local two-species order,
as one can discern from the top left panel of Fig.~\ref{fig42}.
Apparently the competition between the two phases is very
balanced in this small system. It would be interesting to simulate larger
systems and explore whether this type of `phase competition' persists, or
whether the KLS order eventually becomes stable.

Additional information is provided by the susceptibilities. 
Observations of $\Delta \left( 0,1\right) $ are consistent
\begin{figure}[bp]
\begin{center}
\vspace{-.5cm}
\begin{minipage}{\columnwidth}
  \epsfig{file=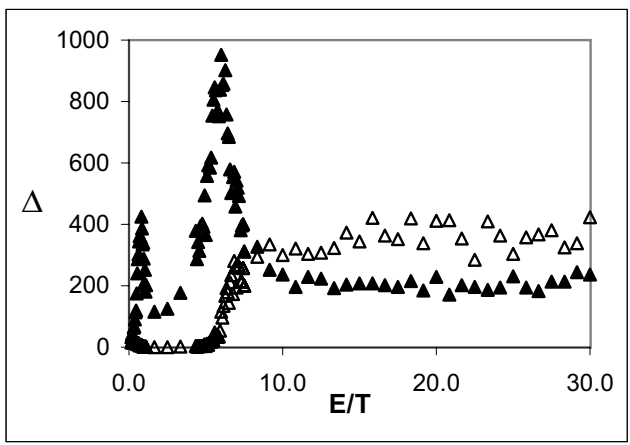, width=\columnwidth}
    \vspace{-.9cm}
\end{minipage}
\end{center}
\caption{$\Delta\left(0,1\right)$ (filled triangles) and 
$\Delta\left(1,0\right)$ (open triangles) as a function of $E/T$ 
for $f=0.075$.} 
\vspace{-.1cm}
\label{fig47}
\end{figure}
with a continuous transition from disorder into the transverse strip at
small $E$. Seeking a signature of the KLS phase, we show $\Delta \left(
1,0\right) $ and $\Delta \left( 0,1\right) $ for $f=0.075$ in Fig.~\ref%
{fig47}. Here we observe a second peak in $\Delta \left( 0,1\right) $ at $%
E=7.2$, which corresponds to $S\left( 0,1\right) \simeq 0.4$ in Fig. \ref%
{fig44}. This peak is rather broad and its amplitude is more than twice that
of the first peak (associated with the DO-TS transition). Perhaps it
suggests a first-order transition which would be observed in a larger
system, separating the two-species phase from the KLS phase. Finally, at
higher $E$ both $\Delta \left( 0,1\right) $ and $\Delta \left( 1,0\right) $
fluctuate around nonzero values, reflecting the fluctuations of the 
competing phases. We note no signal of a transition in 
$\Delta \left( 1,0\right) $.

We now continue by looking at the data for $f=0.05$, shown in Fig.~\ref%
{fig48}, where we have plotted $S$ for each type of order. We observe that
\begin{figure}[tp]
\begin{center}
\vspace{-.1cm}
\begin{minipage}{\columnwidth}
  \epsfig{file=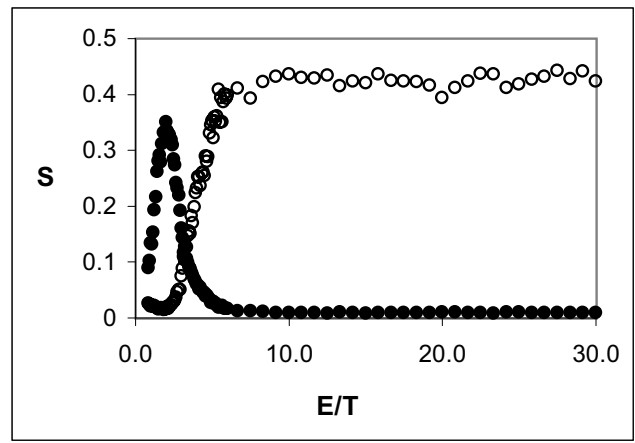, width=\columnwidth}
    \vspace{-.9cm}
\end{minipage}
\end{center}
\caption{$S\left(0,1\right)$ (filled circles) and $S\left(1,0\right)$ (open
circles) as a function of $E/T$ for $f=0.05$.} 
\vspace{-.2cm}
\label{fig48}
\end{figure}
the transverse strip only makes a brief appearance, with $S\left( 0,1\right) 
$ reaching a maximum of only $0.35$. At higher $E$, the parallel strip
stabilizes and reaches a value slightly above $0.4$. 
However, we observe no clear signature of a distinct transition
between the two types of order. There is a vague remnant in $\Delta \left(
0,1\right) $ of the double peak structure seen at $f=0.075$, and then a
broad, messy bump in $\Delta \left( 1,0\right) $ before it settles down to
capture the fluctuations of the ordered phase. Only further investigation of
larger systems can determine whether these weak signals in fact indicate a
transition. It is clear, however, that the high $E$ phase here is not 
the `phase competition' seen previously at $f=0.075$, since neither 
$S\left( 0,1\right) $ nor $\Delta \left(0,1\right) $ contain any trace
of it. 
Upon reducing $f$ below $0.05$ we find that the transverse strip
has essentially disappeared. Meanwhile, $S\left( 1,0\right) $ looks similar
to Fig.~\ref{fig48}, saturating to about $0.50$, indicating that the KLS
strip has formed. Apparently the system is unable to completely order at any 
$E$, since $T_{KLS}\left( E=\infty \right) =1.4$. We stress that this behavior
is due to the proximity of the KLS phase transition, not due to some
residual competition with the transverse strip. Finally, we note that the
behavior of the susceptibility is consistent with a continuous transition
into the parallel strip.

\begin{figure}[bp]
\begin{center}
\vspace{-.1cm}
\begin{minipage}{\columnwidth}
  \epsfig{file=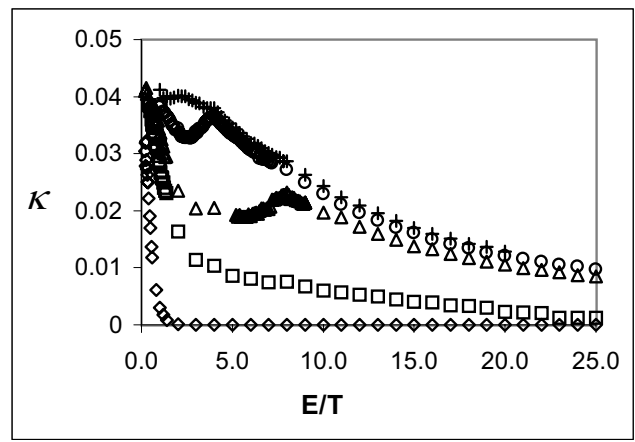, width=\columnwidth}
    \vspace{-.9cm}
\end{minipage}
\end{center}
\caption{Conductivity in the $T=1.2$ plane as a function of $E/T$ for
several $f$. $f=0.50$, diamonds; $f=0.10$, squares; 
$f=0.075$, triangles; $f=0.05$, circles; $f=0.04$, $+$'s.} 
\vspace{-.2cm}
\label{fig52}
\end{figure}

The conductivities are shown in Fig.~\ref{fig52}. At $f=0.50$ and $0.10$, $%
\kappa $ has much the same form as seen before, decreasing monotonically
with $E$. This is no longer the case when $f$ is reduced to $0.075$. Now $%
\kappa $ develops a broad minimum which coincides with the appearance of the
two-species order, followed by a small peak at $E=8$ before falling off with 
$E$. The peak coincides with the crossing of $S\left( 0,1\right) $ and $%
S\left( 1,0\right) $, which suggests that the current is maximized by the
`phase competition'. With the current maximized, increasing $E$ simply
reduces $\kappa $; indeed a glance at the raw data for the current shows
that this is the case. The data for $f=0.05$ shows the same nonmonotonic
structure as for the $f=0.075$ data. Again, the peak corresponds to the
crossing of the two order parameters, and then falls off with increasing $E$.
This similarity suggests that perhaps the high $E$ phase at $f=0.075$ is
in fact the parallel strip, and such behavior would be observed in a larger
system. For values of $f$ below $0.05$ we no longer observe the secondary 
peak, nor do we observe any phase competition. We must, however, leave questions
about the nature of the high-$E$ phase open to further study.

\subsection{Scaling Arguments}

Now that we have surveyed the phase diagram in some detail, we turn to look
closely at some of the boundaries with the help of some scaling arguments.
Considering Fig.~\ref{fig1} again, we note two potentially troubling
features. First, there is a shift in $T_{c}$ across the DO-TS boundary of
about $50\%$ between the two system sizes. We will characterize this shift
using a mean-field scaling argument. Second, we note that the bicritical
point has shifted towards the $f=0$ axis in the larger system. It has been
alluded to before that the number of rows of the minority species, rather
than the fraction $f$, might be the controlling variable. We will
investigate this suggestion more carefully by considering some larger
systems and rectangular geometries. Finally we study the $f=0$ phase
transition, using scaling arguments developed for the KLS model.

The shift in $T_{c}$ with system size is most pronounced at $f=0.50$.
Previous work on the two-species model with $J=0$ treated the ordered phase
in a mean-field approximation by solving equations of motion for the two
different charge densities \cite{LZ, VZS}.\textbf{\ }It was found that the
scaling functions depend on the combination $EL_{y}/T$, indicating that the
effective bias $E/T$ introduces a new length scale. This scaling implies an
infinite-volume limit in which $E/T$ $\rightarrow 0$ as $L_{y}\rightarrow
\infty $, while keeping $EL_{y}/T$ fixed. Earlier analyses of the \emph{%
ordered phase} based on these ideas have worked quite well, so that we now
attempt to extend this approach to analyze quantities near criticality and
\begin{figure}[tp]
\begin{center}
\vspace{-.1cm}
\begin{minipage}{\columnwidth}
  \epsfig{file=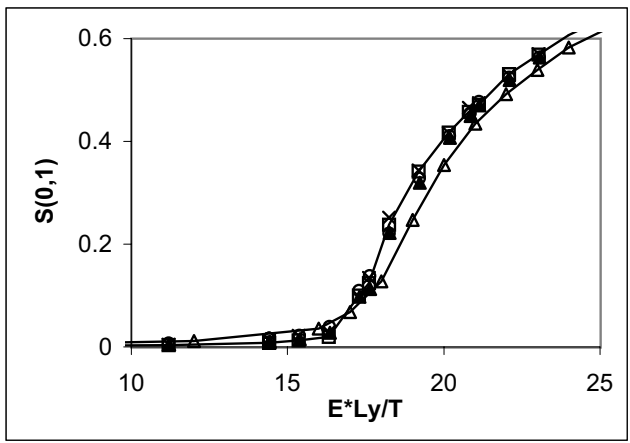, width=\columnwidth}
    \vspace{-.9cm}
\end{minipage}
\end{center}
\caption{$S\left(0,1\right)$ as a function of effective drive for $J=1$.
System sizes: $40\times40$, open triangles; $40\times60$, circles; 
$40\times80$, $\times$'s; $60\times60$, filled triangles; $60\times80$,
squares. Lines have been added to the largest and smallest systems to 
guide the eye.} 
\vspace{-.2cm}
\label{fig31}
\end{figure}
\begin{figure}[bp]
\begin{center}
\vspace{-.5cm}
\begin{minipage}{\columnwidth}
  \epsfig{file=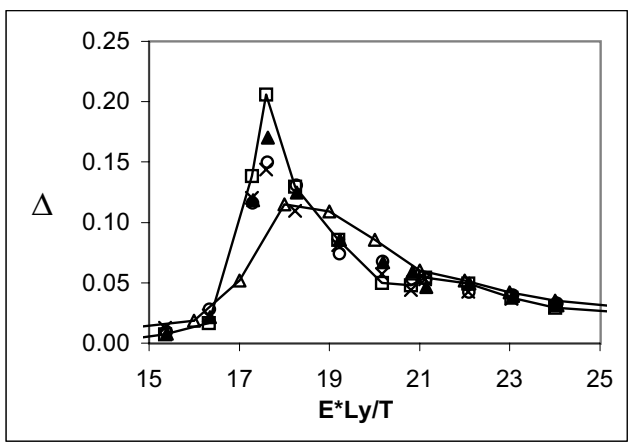, width=\columnwidth}
    \vspace{-.9cm}
\end{minipage}
\end{center}
\caption{$\Delta\left(0,1\right)$ as a function of effective drive for 
$J=1$. The symbols are the same as in the previous figure.} 
\vspace{-.1cm}
\label{fig32}
\end{figure}
for $J\neq 0$. There is no reason to expect success a priori, as both
critical fluctuations and nonzero $J$ may modify the scaling variables and
the mean-field exponents. In Figs.~\ref{fig31} and~\ref{fig32} we have
plotted $S\left( 0,1\right) $ and $\Delta \left( 0,1\right) $ for $J=1.0$.
(We have divided $\Delta \left( 0,1\right) $, Eqn. (\ref{Delta}), by the
volume in order to compare different system sizes more easily.) Rather than
crossing the phase boundary by varying $T$ we have opted instead to vary $E$
since this allows us to vary the effective bias $E/T$ at \emph{constant}
interaction strength, $J/T$ . While the collapse of $S\left( 0,1\right) $ in
Fig.~\ref{fig31} is not perfect, the mean-field scaling argument accounts
for most of the shift in $T_{c}$. There is a shift in the peak of $\Delta
\left( 0,1\right) $ of about $3.6\%$ between the largest and smallest
systems; if the same data were plotted without rescaling, the shift in $T_{c}
$ is about $52\%$. Also of note is the extremely weak dependence on the
transverse dimension, as predicted by the scaling argument. We also examined
the same quantities for $J=0$. Interestingly, the data collapse for $J=1.0$
is better than in the $J=0$ case. This is somewhat surprising, as the
original scaling argument was derived for $J=0$, and we would expect the
interactions to perhaps modify it. Whatever the resolution
of this puzzle, it is apparent that the scaling argument presented here
explains the pronounced shift in $T_{c}$ seen in the phase diagram:
increasing $L_{y}$ requires a corresponding decrease in the effective 
bias $E/T$.

\begin{figure}[bp]
\begin{center}
\vspace{-.5cm}
\begin{minipage}{\columnwidth}
  \epsfig{file=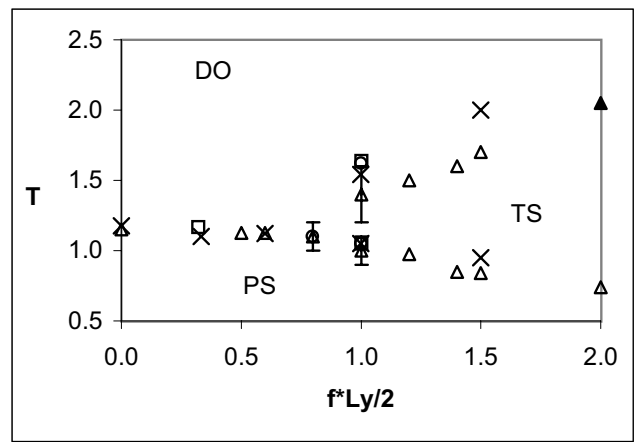, width=\columnwidth}
    \vspace{-.9cm}
\end{minipage}
\end{center}
\caption{The phase diagram in the $f$-$T$ plane, with $f$ rescaled 
to represent the number of rows of the minority species. 
System sizes: filled triangles, $40\times40$; $\times$'s, 
$60\times60$; squares, $60\times80$.} 
\vspace{-.1cm}
\label{fig33}
\end{figure}

Another issue concerns the location of the junction of the three phase
boundaries, shown in Fig.~\ref{fig1}. It is clear that the junction moves
toward the $f=0$ axis as the system size is increased. In fact, in both
systems the boundaries merge just below a \emph{single row} of the minority
species, which naturally corresponds to a smaller $f$ in the larger system.
In Fig.~\ref{fig33} we have replotted the data from Fig.~\ref{fig1},
replacing $f$ with $fL_{y}/2$, which is simply the \emph{number of rows} of
the minority species. Near the junction of the three lines we have also
included results from a few other system sizes with rectangular geometries.
Plotted vs.~$fL_{y}/2$, the junctions of the boundaries coincide, within the
error bars, for all system sizes, suggesting that the onset of the
two-species order occurs, at least in relatively small finite systems, when
there are sufficient minority particles to form a single row. The crucial
question concerns the extrapolation of this result to an appropriate
thermodynamic limit. If the system size goes to infinity in the most naive
way, i.e., $L_{x},L_{y}\rightarrow \infty $ at fixed aspect ratio 
$L_{x}/L_{y}$, the particle density associated with a `single row' vanishes.
It is possible that the DO-PS transition exists in an infinite volume only
at $f=0$, and any \emph{finite density} of `disorder' (i.e., the minority
species) induces the two-species order. Preliminary studies \cite{BSunpub}
indicate that the minority species does indeed constitute a relevant
perturbation to the KLS fixed point. We will have to leave discussion of
this issue to future work and for now limit ourselves to
statements about finite systems.

At $f=0$ there is only one species, and we observe the KLS transition at 
\emph{finite} $E$. Though a great deal of study has been devoted to this
transition at infinite $E$, there has been no detailed work at finite $E$.
Here we present a basic finite-size scaling (FSS)\ analysis of this
transition, in order to locate $T_{c}\left( E=2.0\right) $, and also to
demonstrate the subtleties which can arise when studying phase transitions
with anisotropic, nonequilibrium dynamics.

\begin{figure}[tp]
\begin{center}
\vspace{-.1cm}
\begin{minipage}{\columnwidth}
  \epsfig{file=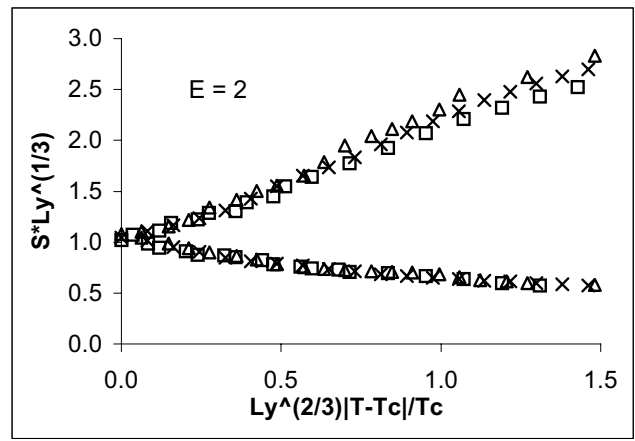, width=\columnwidth}
    \vspace{-.9cm}
\end{minipage}
\end{center}
\caption{Anisotropic scaling plot of $S\left(1,0\right)$ at $f=0$. System
sizes: Squares, $24\times54$; $\times$'s, $28\times86$; triangles, 
$32\times 128$.} 
\vspace{-.2cm}
\label{fig34}
\end{figure}

Field-theoretic studies of the KLS model \cite{FTDDS} indicate that the
critical behavior is strongly anisotropic, meaning that correlation lengths
diverge with different exponents in the field direction and perpendicular to
the field. Specifically, the fluctuations perpendicular to the field are
gaussian $\left( \nu _{\bot }=1/2\right) $ while those parallel to the field
are not $\left( \nu _{\Vert }=3/2\right) $. Correlations therefore grow
faster in the parallel direction as $T\rightarrow T_{c}(E)$, suggesting an
analysis of rectangular samples such that the anisotropic aspect ratio $%
A\equiv L_{\Vert }^{\nu _{\bot }/\nu _{\Vert }}L_{\bot }^{-1}$ is held fixed %
\cite{LeungDDS}. While there is some discussion regarding the correct
mesoscopic model \cite{debate}, detailed numerical simulations show that the
exponents cited above are the correct ones \cite{WangDDS, ItalDDS}. In the
following we will use only the phenomenological result of Leung for the
scaling of the order parameter at fixed $A$: 
\begin{equation}
S\left( T,L_{\Vert },L_{\bot }\right) =L_{\Vert }^{-\beta /\nu _{\Vert }}%
\overline{S}\left( tL_{\Vert }^{1/\nu _{\Vert }},L_{\Vert }^{\nu _{\bot
}/\nu _{\Vert }}L_{\bot }^{-1}\right)
\end{equation}%
%
where $S$ refers to $S\left( 1,0\right) $. A detailed discussion of the
subtleties of the FSS analysis for the KLS model and precision numerical
results can be found in \cite{ItalDDS}. Fig.~\ref{fig34} presents our data
for the scaled order parameter at $E=2$; the same data for saturation $E$
can be found elsewhere \cite{lymanCPS}. This data is not intended as a test
of the mesoscopic model, it merely is meant to indicate that the exponents
at infinite $E$ are consistent with those at finite $E$, and to determine $%
T_{c}\left( E=2.0\right) =1.20\left( 2\right) $. The data collapse is
comparable to that seen at saturation $E$, with the high temperature (upper)
branch collapsing quite well and the low temperature (lower) branch showing
small, but systematic deviations from scaling. These deviations 
remain unexplained. They are possibly due to corrections to scaling or 
perhaps the asymptotic region is only observed very close to $T_{c}$.

We have seen how subtle are the issues surrounding the transition at $f=0$.
Perhaps a fruitful way to proceed when $f\neq 0$ is to adopt the technique
introduced by Caracciolo \emph{et al}. \cite{finitepsi}, directly measuring
finite volume correlation lengths for various geometries and volumes. In
this way we may develop some understanding of how to approach the infinite
volume limit in a simple way, minimizing corrections to scaling which would
complicate an uninformed analysis.

\section{Conclusions and Outlook}

We have compiled a detailed phase diagram for a system of two species of
particles, interacting via attractive Ising interactions, and driven into
opposite directions by an external ``electic'' field, $E$. The purpose of
our analysis was to unify previous studies which were restricted either to
just one species or to having only excluded volume interactions. In the
former case, particles order into a single strip aligned with the field
direction while in the second model, the two oppositely driven species form
a jam in the shape of a transverse strip. We monitor structure factors,
their fluctuations, and, if necessary, their time traces to identify the
location and character of the transitions. Most of our data are taken at
fixed $E$, varying the fraction $f$ of the `minority' species and
temperature $T$. At high $f$, we observe a continuous transition from a
disordered phase into the transverse strip, as the temperature is lowered.
Noting a significant system-size dependence of the critical line, we invoke
a mean-field scaling argument \cite{LZ, VZS} which suggests that $EL_{y}/T$
is a good scaling variable. This is confirmed quite satisfactorily by our
data. At smaller $f$ ($0.05\leq f\leq 0.10$ for a $40\times 40$ system) we
observe two transition as $T$ is lowered: first into the transverse strip
(continuous), and then into the parallel strip (first order). And finally at
the smallest $f$ a single, continuous transition is observed into the
parallel strip. The junction of the three phases -- disorder, transverse and
parallel strip -- appears to be a multicritical point. Analyzing data for a
range of system sizes suggests that its location scales with $f/L_{x}$,
rather than $f$; i.e., the relevant quantity is the number of \emph{rows}
which can be formed by the minority particles. Depending on how
the thermodynamic limit is approached, the multicritical point can shift  
to $f=0$. 

Several projects suggest themselves to extend this work. First, an analysis
of larger systems at fixed $T$ could clear up some questions, especially
regarding the fate of the `phase competition' which is observed close to the
multicritical point. Many of the `transitions' in the $40\times 40$ system
require an analysis of larger lattices in order to confirm their existence.
Second, a look at structure factors in the disordered phase is likely to
reveal the presence of long-range correlations. Since these are known to be
quite distinct in the KLS \cite{eff-LR} and SHZ \cite{2sp-corr} models, it
would be interesting to investigate how the crossover occurs. Such a study
would also yield considerable insight into the type of noise terms which
would have to be added to the mean-field equations in order to capture
fluctuations and critical properties accurately. These equations would then
provide a reliable starting point for an analysis of the KLS transition in
the presence of a few minority charges, in order to understand the true
nature of the KLS critical point: does it mark the beginning of a critical
line, or is it a multicritical point?

\textbf{Acknowledgements. }We thank R.K.P.~Zia, I.~Georgiev, U.C.~T\"{a}%
uber, A.~Gambassi, M.~Gubinelli, G.~Korniss, and H.K.~Janssen for fruitful
discussions. This work was supported in part by NSF grants DMR-0088451, 
DMR-0414122, and SBE-0244916, as well as the Jeffress Memorial Trust, 
Grant No. J-594.

\end{document}